\documentclass[sigconf]{acmart}

\usepackage{xcolor}
\usepackage{cleveref}

\acmConference[ICSE 2024]{46th International Conference on Software Engineering}{April 2024}{Lisbon, Portugal}

\AtBeginDocument{%
  \providecommand\BibTeX{{%
    \normalfont B\kern-0.5em{\scshape i\kern-0.25em b}\kern-0.8em\TeX}}}

\copyrightyear{2024}
\acmYear{2024}
\setcopyright{acmlicensed}\acmConference[IDE '24]{2024 First IDE Workshop}{April 20, 2024}{Lisbon, Portugal}
\acmBooktitle{2024 First IDE Workshop (IDE '24), April 20, 2024, Lisbon, Portugal}
\acmDOI{10.1145/3643796.3648451}
\acmISBN{979-8-4007-0580-9/24/04}

\begin{document}

\title{Tool-Augmented LLMs as a Universal Interface for IDEs}

\author{Yaroslav Zharov}
\authornote{Authors contributed equally to this research.}
\email{yaroslav.zharov@jetbrains.com}
\affiliation{%
  \institution{JetBrains Research}
  \city{Karlsruhe}
  \country{Germany}
}

\author{Yury Khudyakov}
\authornotemark[1]
\email{yury.khudyakov@jetbrains.com}
\affiliation{%
  \institution{JetBrains Research}
  \city{Bremen}
  \country{Germany}
}

\author{Evgeniia Fedotova}
\authornotemark[1]
\email{evgeniia.fedotova@jetbrains.com}
\affiliation{%
  \institution{JetBrains Research}
  \city{Belgrade}
  \country{Serbia}
}

\author{Evgeny Grigorenko}
\authornotemark[1]
\email{evgeny.grigorenko@jetbrains.com}
\affiliation{%
  \institution{JetBrains Research}
  \city{Belgrade}
  \country{Serbia}
}

\author{Egor Bogomolov}
\authornotemark[1]
\email{egor.bogomolov@jetbrains.com}
\affiliation{%
  \institution{JetBrains Research}
  \city{Amsterdam}
  \country{the Netherlands}
}








\renewcommand{\shortauthors}{Zharov et al.}

\begin{abstract}
  Modern-day Integrated Development Environments (IDEs) have come a long way from the early text editing utilities to the complex programs encompassing thousands of functions to help developers.
  However, with the increasing number of efficiency-enhancing tools incorporated, IDEs gradually became sophisticated software with a steep learning curve.
  The rise of the Large Language Models (LLMs) capable of both natural language dialogue and code generation leads to a discourse on the obsolescence of the concept of IDE.
  In this work, we offer a view on the place of the LLMs in the IDEs as the universal interface wrapping the IDE facilities.
  We envision a model that is able to perform complex actions involving multiple IDE features upon user command, stripping the user experience of the tedious work involved in searching through options and actions.
  For the practical part of the work, we engage with the works exploring the ability of LLMs to call for external tools to expedite a given task execution.
  We showcase a proof-of-concept of such a tool.
\end{abstract}





\maketitle

\section{Outline}

\label{sec:outlines}
With the worldwide position of the IT sector in the economy, it is only natural, that the tools used by the bedrock of this sector--software developers--would grow more sophisticated and specialized.
This growth, however, comes at the price of a steeper learning curve.
To utilize modern development environments at full capacity, a user may need to watch some introductory video and continue constantly sifting through documentation for a long period of alignment.
Not only does it require constant effort even from vetted users, but the cluttered interface, which is a byproduct of the feature development, affects IDE users' productivity, as shown in \cite{Kasatskii23}. Moreover, it becomes a problem for the IDE developers: with the growth in the number of features, it becomes harder to add and promote new functionality.

To improve the discoverability of features, multiple IDEs incorporated the idea of a centralized search box for any type of activity (e.g., Search Everywhere by IntelliJ IDEA or Command Palette by VSCode).
This element provides a universal place to look up all possible preferences, actions, plug-ins, and more, which helps to lift some of the burden. While it aids users, they still need to decompose their tasks and find out the most appropriate way to execute them within the IDE. 

In this short essay, we suggest that tool-augmented language models~\cite{ToolFormer,Gorilla,ToolLLM} can aid both IDE users and IDE developers in their needs. Such models in other domains, in addition to chatting with the user, can make calls to external tools such as web search, calculator~\cite{ToolFormer}, other models~\cite{HuggingGPT,Chameleon}, or an open set of APIs~\cite{ToolLLM}. It allows them to perform more complex scenarios compared to regular language models; scenarios that require planning, precise operations, and harvesting information from external sources. 

We suggest making the internals of the IDEs available to the tool-augmented models as APIs.
This will equip the model with more precise methods to manipulate, execute, and navigate code and other project-related information such as data from version control systems, since the model is forced to explicitly select APIs to call, instead of working on implicit assumption that the right tool can be retrieved by e.g., nearest embedding search.
The proposed modification will grant users an alternative way of IDE interaction, text interface of a model which in turn will communicate with the IDE to manipulate project and answer user queries.


\section{IDE-Augmented LLM-Based Agents}

\subsection{Related Works}

The research in the direction of tool-augmented language models was kick-started by the ToolFormer work~\cite{ToolFormer}, where a model was trained to call one of several tools (e.g., calculator or Wikipedia search) to improve question answering performance.
The Gorilla model~\cite{Gorilla} extended the idea with a tool retrieval submodule, which allowed the execution of the model with an open set of tools.
The ToolLLM model~\cite{ToolLLM} combined this with the~\cite{TreeofThoughts} to create a Depth-First Action Tree Search algorithm.
This improvement allowed the model to perform complex actions and correct itself on the go when the initial plan goes wrong.

On the other hand, there is a plethora of works, working on providing LLMs with more agency to perform complex actions.
Some, like Chameleon~\cite{Chameleon}, employ techniques alike to Chain-of-Thought~\cite{Chain-of-Thought}, first planning the whole execution pipeline and then executing it.
Others, like the aforementioned ToolLLM model~\cite{ToolLLM} or ToolChain*~\cite{Toolchain-star}, employ more complex planning algorithms.
Using an LLM as an agent demonstrates successful cases, like the Voyager model~\cite{Voyager} playing the Minecraft game or ChemCrow~\cite{ChemCrow} solving practical chemistry tasks.

\subsection{Implications for IDE Users}

\begin{figure}
    \centering
    \includegraphics[width=\linewidth]{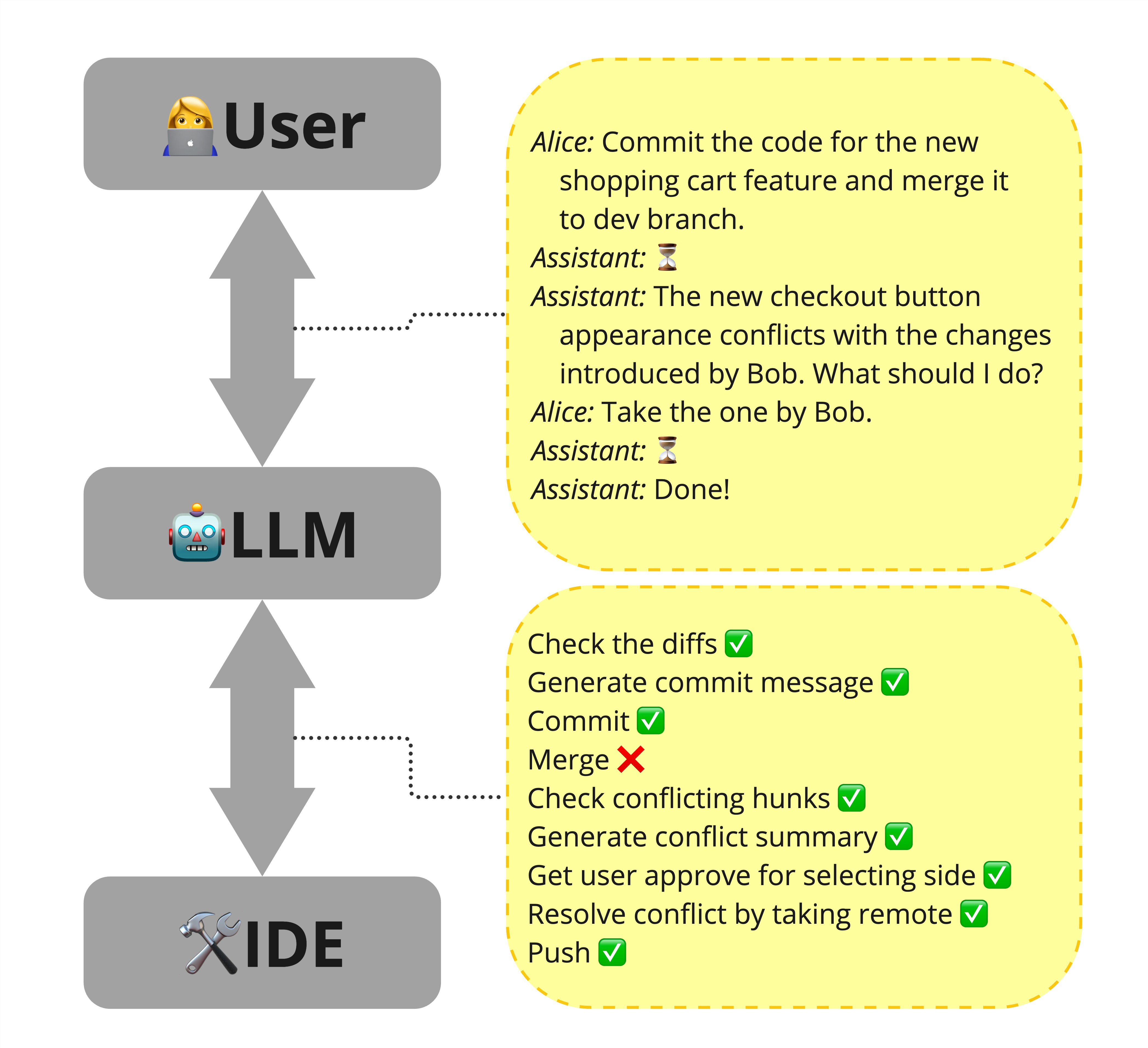}
    \caption{Hypothetical scenario involving CVS merge conflict resolution utilizing interaction with IDE tools.}
    \label{fig:merge-conflict}
\end{figure}

We do not think that the idea to completely replace the interface with an assistant is a viable idea. However, we suppose that this assistant can help users of an IDE in two main types of scenarios: repetitive work and rarely occurring tasks requiring a complicated combination of tools.

An example of the first scenario is the task of VCS conflict resolution (\Cref{fig:merge-conflict}).
While the task itself consists of actions of a simple nature, it is cognitively demanding.
The repetitiveness of such tasks and actions leads to attentional strain.
An example of the second scenario is creating a new project.
While modern-day IDEs are equipped with lots of tools to help set up environments, repositories, directory structures, and other configurations, the task occurs rarely, and therefore tools aren't in the muscular memory of a developer.

\subsection{Implications for IDE Developers}

As we mentioned previously, IDE developers face the problem of discoverability for newly developed features as well as the problem of overloading the IDE interface. With the introduction of the text-based interface where language models act as an intermediary between the user and IDE, IDE developers can focus on making new features accessible for the models (e.g., by writing appropriate documentation). In contrast to the users who have preferences and require teaching, models can instantly benefit from the new functionality and immediately start using it to assist the users. Additionally, if part of the developed functionality is masked from users and only made available for models, the default user interface can be simplified and become more lightweight.

\section{Research Challenges}

Even though the approach itself is similar to the previously presented research, the practical application imposes several research questions.
In this section, we will outline and discuss them briefly.

\paragraph{How complicated is the typical user scenario?}
The current generation of models has a limited capacity to build long chains of tool calls.
Not only due to the limited attention span but also due to the increased reasoning mistakes on long chains~\cite{LengthGeneralization}.
It is important to evaluate the performance of the models on a representative dataset.
Such user requests and statistics being collected in vivo can be an insightful result on its own.

\paragraph{Which model optimally integrates reasoning with software engineering knowledge?}
In order to act as an intermediary between the user and the IDE, models should know all the nuances of the software development process. 
We preliminary tested the base model provided with the ToolLLM package~\cite{ToolLLM}, and noted a prominent lack of ability to understand SE-related requests.
Hypothetically, this is caused by the model being fine-tuned from the Llama 2~\cite{LLAMA2} checkpoint, which is not specifically trained for software engineering.
Our qualitative assessment of other models' performance leads to the conclusion that CodeLlama 2 performs better, which supports the hypothesis that we need a model fluent in SE.
However, we face the question of how different pre-training datasets affect the model's ability to reason and plan on one hand and understand SE-related requests on the other.

\paragraph{Should tool retrieval be yet another tool in the shed?}
In case of the open-set tool usage it is impossible to tune the model to remember all APIs, and due to the vast set of tools it's also impossible to fit the whole set of tools in the context window.
To solve this, referenced approaches, filter the APIs based on the user request, so that the resulting toolset fits in the context window.
While this approach works for relatively homogenous tools explored in those works, it may fail with user requests involving tools from drastically different domains (e.g., both version control and code refactoring).
Potential solution for this can be to implicitly call retrieval tool for each generation step~\cite{AntonDenis}, however this requires model tuning.
We hypothesize that granting the model an explicit access to the tool search can improve it performance by allowing ad-hoc search requests tailored for the local situation in the middle of the plan execution.

\paragraph{Can LLMs be considerate of the environment?}
The approaches listed in the outlines section restricted the model output to a specific format and a specific set of tools.
This requires bringing all APIs to a common interface and--especially for the DFDS~\cite{ToolLLM}--the ability to backtrack the changes introduced by a tool call.
However, this limits the applicability of the method, since every API either shouldn't introduce any changes to the environment, or should be reversible.
We want to assess if the modern-day LLMs have enough awareness to either avoid breaking actions before they are absolutely necessary or revert them on their on.
Models possessing such awareness can be granted access to the full codebase of the IDE to perform appropriate actions in an unrestricted environment, the same way we grant an experienced user higher access to the system administration.

\paragraph{Are models able to benefit from asking clarifying questions? How much?}
It is beneficial for a model to question the user for additional information~\cite{CLAM}.
However, for the commercial use, it's important to finda a balance between the benefits of additional information and the additional compute and user inconvenience.
Measuring the effect of such questions precisely is a research task on its own and requires a thorough dataset collection.
Resolving such a question may impact the way AI assistants interact with users.
On one hand LLMs prefer to hallucinate a prediction instead of asking additional question, on the other after limiting the options to a small number of most probable it may be beneficial to generate all of them and present the user with several variants.
~\\

We believe that if the community of practitioners and researchers answers the outlined challenges, tool-augmented language models can pave a novel way of interacting with the IDEs and make them more transparent and convenient for their users.

\bibliographystyle{ACM-Reference-Format}
\bibliography{references}

\appendix









\end{document}